


\font\bigbold=cmbx12
\font\ninerm=cmr9      \font\eightrm=cmr8    \font\sixrm=cmr6
\font\fiverm=cmr5
\font\ninebf=cmbx9     \font\eightbf=cmbx8   \font\sixbf=cmbx6
\font\fivebf=cmbx5
\font\ninei=cmmi9      \skewchar\ninei='177  \font\eighti=cmmi8
\skewchar\eighti='177  \font\sixi=cmmi6      \skewchar\sixi='177
\font\fivei=cmmi5
\font\ninesy=cmsy9     \skewchar\ninesy='60  \font\eightsy=cmsy8
\skewchar\eightsy='60  \font\sixsy=cmsy6     \skewchar\sixsy='60
\font\fivesy=cmsy5     \font\nineit=cmti9    \font\eightit=cmti8
\font\ninesl=cmsl9     \font\eightsl=cmsl8
\font\ninett=cmtt9     \font\eighttt=cmtt8
\font\tenfrak=eufm10   \font\ninefrak=eufm9  \font\eightfrak=eufm8
\font\sevenfrak=eufm7  \font\fivefrak=eufm5
\font\tenbb=msbm10     \font\ninebb=msbm9    \font\eightbb=msbm8
\font\sevenbb=msbm7    \font\fivebb=msbm5
\font\tenssf=cmss10    \font\ninessf=cmss9   \font\eightssf=cmss8
\font\tensmc=cmcsc10


\newfam\bbfam   \textfont\bbfam=\tenbb \scriptfont\bbfam=\sevenbb
\scriptscriptfont\bbfam=\fivebb  \def\Bbb{\fam\bbfam}
\newfam\frakfam  \textfont\frakfam=\tenfrak \scriptfont\frakfam=%
\sevenfrak \scriptscriptfont\frakfam=\fivefrak  \def\frak{\fam\frakfam}
\newfam\ssffam  \textfont\ssffam=\tenssf \scriptfont\ssffam=\ninessf
\scriptscriptfont\ssffam=\eightssf  
\def\smc{\tensmc}

\def\eightpoint{\textfont0=\eightrm \scriptfont0=\sixrm
\scriptscriptfont0=\fiverm  \def\rm{\fam0\eightrm}%
\textfont1=\eighti \scriptfont1=\sixi \scriptscriptfont1=\fivei
\def\oldstyle{\fam1\eighti}\textfont2=\eightsy
\scriptfont2=\sixsy \scriptscriptfont2=\fivesy
\textfont\itfam=\eightit         \def\it{\fam\itfam\eightit}%
\textfont\slfam=\eightsl         \def\sl{\fam\slfam\eightsl}%
\textfont\ttfam=\eighttt         \def\tt{\fam\ttfam\eighttt}%
\textfont\frakfam=\eightfrak     \def\frak{\fam\frakfam\eightfrak}%
\textfont\bbfam=\eightbb         \def\Bbb{\fam\bbfam\eightbb}%
\textfont\bffam=\eightbf         \scriptfont\bffam=\sixbf
\scriptscriptfont\bffam=\fivebf  \def\bf{\fam\bffam\eightbf}%
\abovedisplayskip=9pt plus 2pt minus 6pt   \belowdisplayskip=%
\abovedisplayskip  \abovedisplayshortskip=0pt plus 2pt
\belowdisplayshortskip=5pt plus2pt minus 3pt  \smallskipamount=%
2pt plus 1pt minus 1pt  \medskipamount=4pt plus 2pt minus 2pt
\bigskipamount=9pt plus4pt minus 4pt  \setbox\strutbox=%
\hbox{\vrule height 7pt depth 2pt width 0pt}%
\normalbaselineskip=9pt \normalbaselines \rm}

\def\ninepoint{\textfont0=\ninerm \scriptfont0=\sixrm
\scriptscriptfont0=\fiverm  \def\rm{\fam0\ninerm}\textfont1=\ninei
\scriptfont1=\sixi \scriptscriptfont1=\fivei \def\oldstyle%
{\fam1\ninei}\textfont2=\ninesy \scriptfont2=\sixsy
\scriptscriptfont2=\fivesy
\textfont\itfam=\nineit          \def\it{\fam\itfam\nineit}%
\textfont\slfam=\ninesl          \def\sl{\fam\slfam\ninesl}%
\textfont\ttfam=\ninett          \def\tt{\fam\ttfam\ninett}%
\textfont\frakfam=\ninefrak      \def\frak{\fam\frakfam\ninefrak}%
\textfont\bbfam=\ninebb          \def\Bbb{\fam\bbfam\ninebb}%
\textfont\bffam=\ninebf          \scriptfont\bffam=\sixbf
\scriptscriptfont\bffam=\fivebf  \def\bf{\fam\bffam\ninebf}%
\abovedisplayskip=10pt plus 2pt minus 6pt \belowdisplayskip=%
\abovedisplayskip  \abovedisplayshortskip=0pt plus 2pt
\belowdisplayshortskip=5pt plus2pt minus 3pt  \smallskipamount=%
2pt plus 1pt minus 1pt  \medskipamount=4pt plus 2pt minus 2pt
\bigskipamount=10pt plus4pt minus 4pt  \setbox\strutbox=%
\hbox{\vrule height 7pt depth 2pt width 0pt}%
\normalbaselineskip=10pt \normalbaselines \rm}

\global\newcount\secno \global\secno=0 \global\newcount\meqno
\global\meqno=1 \global\newcount\appno \global\appno=0
\newwrite\eqmac \def\romappno{\ifcase\appno\or A\or B\or C\or D\or
E\or F\or G\or H\or I\or J\or K\or L\or M\or N\or O\or P\or Q\or
R\or S\or T\or U\or V\or W\or X\or Y\or Z\fi}
\def\eqn#1{ \ifnum\secno>0 \eqno(\the\secno.\the\meqno)
\xdef#1{\the\secno.\the\meqno} \else\ifnum\appno>0
\eqno({\rm\romappno}.\the\meqno)\xdef#1{{\rm\romappno}.\the\meqno}
\else \eqno(\the\meqno)\xdef#1{\the\meqno} \fi \fi
\global\advance\meqno by1 }

\global\newcount\refno \global\refno=1 \newwrite\reffile
\newwrite\refmac \newlinechar=`\^^J \def\ref#1#2%
{\the\refno\nref#1{#2}} \def\nref#1#2{\xdef#1{\the\refno}
\ifnum\refno=1\immediate\openout\reffile=refs.tmp\fi
\immediate\write\reffile{\noexpand\item{[\noexpand#1]\ }#2\noexpand%
\nobreak.} \immediate\write\refmac{\def\noexpand#1{\the\refno}}
\global\advance\refno by1} \def\semi{;\hfil\noexpand\break ^^J}
\def\nl{\hfil\noexpand\break ^^J} \def\refn#1#2{\nref#1{#2}}

\def\prl#1#2#3{{\it Phys.\ Rev.\ Lett.}\ {\bf #1} ({#2}) #3}

\newif\iftitlepage \titlepagetrue \newtoks\titlepagefoot
\titlepagefoot={\hfil} \newtoks\otherpagesfoot \otherpagesfoot=%
{\hfil\tenrm\folio\hfil} \footline={\iftitlepage\the\titlepagefoot%
\global\titlepagefalse \else\the\otherpagesfoot\fi}

\def\abstract#1{{\parindent=30pt\narrower\noindent\ninepoint\openup
2pt #1\par}}

\newcount\notenumber\notenumber=1 \def\note#1
{\unskip\footnote{$^{\the\notenumber}$} {\eightpoint\openup 1pt #1}
\global\advance\notenumber by 1}

\def\today{\ifcase\month\or January\or February\or March\or
April\or May\or June\or July\or August\or September\or October\or
November\or December\fi \space\number\day, \number\year}

\def\pagewidth#1{\hsize= #1}  \def\pageheight#1{\vsize= #1}
\def\hcorrection#1{\advance\hoffset by #1}
\def\vcorrection#1{\advance\voffset by #1}

\pageheight{23cm}
\pagewidth{15.7cm}
\hcorrection{-1mm}
\magnification= \magstep1
\parskip=5pt plus 1pt minus 1pt
\tolerance 8000
\def\bsk{\baselineskip= 14.5pt plus 1pt minus 1pt}
\bsk

\font\extra=cmss10 scaled \magstep0  \setbox1 = \hbox{{{\extra R}}}
\setbox2 = \hbox{{{\extra I}}}       \setbox3 = \hbox{{{\extra C}}}
\setbox4 = \hbox{{{\extra Z}}}       \setbox5 = \hbox{{{\extra N}}}





\def\frac#1#2{{#1\over#2}}

\def\pmb#1{\setbox0=\hbox{$#1$} \kern-.025em\copy0\kern-\wd0
    \kern.05em\copy0\kern-\wd0 \kern-.025em\raise.0433em\box0 }

\def\ve{\vfill\eject}

\def\frac#1#2{{#1\over#2}}

\def\pmb#1{\setbox0=\hbox{$#1$} \kern-.025em\copy0\kern-\wd0
    \kern.05em\copy0\kern-\wd0 \kern-.025em\raise.0433em\box0 }

\def\ve{\vfill\eject}

\def\({\left(}
\def\){\right)}

\def\omit#1{}    
 
  \def\CB{\copy1} \def\RB#1{\raise#1\CB}
  \def\CC{\copy2} \def\RC#1{\raise#1\CC}
  \def\xex#1{\x#1ex}       \def\yex#1{\y#1ex}    
         
\def\x{\hskip}              
\def\y{\vskip}              
             \def\bitt{\xex{.3}}    
\def\bit{\xex{.15}}      
  \def\bittt{\xex{.45}}  
\def\m#1{$#1$}       
\def\mm#1{$\,#1\,$}   
   \def\newline{\hfill\break}

\def\m#1{$#1$} \def\mm#1{$\,#1\,$} 

\def\E#1#2{$$ #2 \eqn{#1} $$}
\def\f#1#2{{#1\over#2}}     
\def\ff#1#2{\raise.5pt\hbox{\eightpoint${\displaystyle\f{#1}{#2}}$}}



          

\def\co{\, ,}   \def\pe{\, .}   


\def\ordo#1{ {\cal O} \( #1 \) }
\def\s#1{\sqrt{#1}}
\def\lt{\left}    \def\({\lt(}    \def\[{\lt[}    \def\<{\lt\langle}
\def\rt{\right}   \def\){\rt)}    \def\]{\rt]}    \def\>{\rt\rangle}
\def\_{^{}_}

\def\intl{\int\limits}   \def\suml{\sum\limits}


\def\text#1{{   \rm     #1    }}     
     
\def\texttt#1{{ \rm     #1 \  }}     
     
\def\t#1{{\rm #1}}  
\def\text#1{\t{#1}}        
  
\def\texttt#1{\t{#1}\ }
\def\d{\text{d}}
\def\ee#1{{\text{e}^{#1}}}


\def\qf{y}

\def\qk{k}

\def\qm{m}
\def\qq{q}
\def\qs{\tau}
\def\qt{t}

\def\qA{A}

\def\qN{N}
\def\qsigma{\sigma}

\def\Dir{^-}
\def\Neu{^+}

\def\figone{qfllow5.eps}
\def\figtwo{qflmed7.eps}
\def\figthree{qflhigh6.eps}

\input epsf

\let\omitpictures=N


{

\refn\CFT
{T. Cheon, T. F\"{u}l\"{o}p and I. Tsutsui,
{\it Ann.\ Phys.} {\bf 294} (2001) 1}

\refn\Warburton
{R.J. Warburton, C. Sch\"aflein, D. Haft, F. Bickel, A. Lorke, K. Karrai, J.M.
Garcia, W. Scoenfeld and P.M. Petroff, {\it Nature} {\bf 405} (2000) 926}

\refn\Lorke
{A. Lorke, R.J. Luyken, A.O. Govorov, J.P. Kotthaus, J.M. Garcia and P.M.
Petroff, \prl{84}{2000}{2223}}

\refn\Nature
{A. Fuhrer, S. L{\"u}sher, T. Ihn, T. Heinzel, K. Ensslin, W. Wegscheider
and M. Bichler, {\it Nature} {\bf 413} (2001) 822}

\refn\FTC
{T. F\"{u}l\"{o}p,  I. Tsutsui and T. Cheon,
{\sl Spectral Properties on a Circle with a Singularity},
{\it J.\ Phys.\ Soc.\ Japan}, to appear; quant-ph/0307002}

\refn\elsewhere
{to be reported elsewhere}

}



\pageheight{23cm}
\pagewidth{15.7cm}
\hcorrection{0mm}
\magnification= \magstep1
\def\bsk{%
\baselineskip= 16.8pt plus 1pt minus 1pt}
\parskip=5pt plus 1pt minus 1pt
\tolerance 6000




\hfill 
\vskip -4pt 
\hfill 
\phantom{quant-ph/0207xxx}

\vskip 42pt

{\baselineskip=18pt

\centerline{\bigbold
Quantum Force due to Distinct Boundary Conditions}

\vskip 30pt

\centerline{\smc
Tam\'{a}s F\"{u}l\"{o}p\footnote{${}^{*}$}
{\eightpoint email:\ fulopt@post.kek.jp},
\quad
Hitoshi Miyazaki
\quad
{\rm and}
\quad
Izumi Tsutsui\footnote{${}^\dagger$}
{\eightpoint email:\ izumi.tsutsui@kek.jp}
}

\vskip 15pt

{
\baselineskip=13pt
\centerline{\it
Institute of Particle and Nuclear Studies}
\centerline{\it
High Energy Accelerator Research Organization (KEK)}
\centerline{\it Tsukuba 305-0801}
\centerline{\it Japan}
}

\vskip 100pt

\abstract{%
{\bf Abstract.}\quad
We calculate the quantum statistical force acting on a partition wall that divides a one
dimensional box into two halves. The two half boxes contain the same (fixed) number of
noninteracting bosons, are kept at the same temperature, and admit the same boundary
conditions at the outer walls; the only difference is the distinct
boundary conditions imposed at the two sides of the partition wall.
The net force acting on the partition wall is nonzero at zero temperature and remains almost
constant for low temperatures. As the temperature increases, the force starts to decrease
considerably, but after reaching a minimum it starts to increase, and tends to infinity
with a square-root-of-temperature asympotics. This example demonstrates clearly that distinct
boundary conditions cause remarkable physical effects for quantum systems.}

\vskip 10pt
%
%
%
%
}
\ve


\pageheight{23cm}
\pagewidth{15.7cm}
\hcorrection{-1mm}
\magnification= \magstep1
\def\bsk{%
\baselineskip= 15.2pt plus 1pt minus 1pt}
\parskip=5pt plus 1pt minus 1pt
\tolerance 8000
\bsk


\secno=1 \meqno=1

\bigskip
\noindent{\bf 1. Introduction}
\medskip

Quantum systems in less than three dimensions enjoy an increasing popularity
and importance in physics. These seemingly simple systems (rings, boxes, ``dots", etc.)
exhibit unexpectedly interesting properties [\CFT], many of which have originally been
found in connection with quantum field theories. The recent developments in nanotechnology
now make it possible to manufacture such quantum mechanical devices, allowing us to
study these various features in laboratories
[\Warburton--\Nature].
With tunable control parameters --- such as varying the magnetic flux that is driven through
a ring, or influencing the parameters that characterize the boundary conditions at the walls
of a box --- these properties become governable, giving us thus controllable quantum devices
that may be useful for many technological applications.

In fact, the dependence of the physical properties on the control parameters is quite
strong. For example, the energy spectrum of a box or a ring with junctions changes
considerably with different boundary/fitting conditions applied at the walls/junctions
[\FTC]. The aim of the present paper is to demonstrate a physical consequence of this 
effect in the context of quantum statistical mechanics. Namely, we show by the example of the
box systems how different quantum statistical behavior emerges from distinct boundary conditions
and the corresponding distinct spectra. One of the remarkable results is that not only the
low-temperature behavior is sensitive to the difference in spectrum but also the
high-temperature properties depend clearly on it.

The setting we consider is a one-dimensional quantum well/box, divided into two halves
with an internal partition wall. Both halves contain the same number of
noninteracting particles. We impose the same (Dirichlet) boundary conditions at the two ends
of the box, while the boundary conditions at the two sides of the partition wall are
chosen to be different (Dirichlet from the left, Neumann from the right). {}From the different
emerging spectra in the two half regions, we calculate the quantum statistical force/pressure
acting on the internal wall from the left and from the right, and the corresponding net force
acting on it, as the function of temperature. We present both numerical results (which are
obtained by appropriate truncations of the arising infinite sums) and analytical
approximations, the latter ones aiming at understanding the low and the high temperature
regimes.

The number of particles on both sides, \m{N}, is arbitrary, and is not necessarily
macroscopically large. Our numerical results are presented for \mm{N = 100}, which is a
realistic population number in nanoscale quantum experiments [\Nature]. The particles are
considered as bosons in this paper, however, we mention that the results prove to be
qualitatively (and partly quantitatively) similar for fermionic particles, too [\elsewhere].

\secno=2 \meqno=1

\bigskip
\noindent{\bf 2. Low temperature regime}
\medskip

The system we consider is formulated as a one-dimensional quantum well of width $2l$ given 
by the interval $[-l, l]$ with a  partition placed at the centre $x = 0$. At the
end walls of the well the states are supposed to obey the
Dirichlet boundary condition $\psi(\pm l) = 0$. At the centre, we assume that the
partition imposes distinct boundary conditions on the left and the right, our choice is
to impose the Dirichlet one for $ x = -0$ and the Neumann one for $x = +0$: 
$$
\psi(-0) = 0, \qquad
\psi'(+0) = 0.
\eqn\no
$$
The two half wells seperated by the partition then admit
the energy levels $E^{\pm}_n = e^{\pm}_n {\cal E}$, $n = 1$, 2, 3 $\ldots$, with 
$$
e^{+}_n = \left(n-{1\over 2}\right)^2, \qquad
e^{-}_n = n^2, \qquad 
{\cal E} = {{\hbar^2}\over{2m}}\left({{\pi}\over{l}}\right)^2.
\eqn\unitenergy
$$

Suppose that we put $N$ identical bosonic particles into each of the two half
wells.  The particles will then distribute among the eigenstates according to the
Bose-Einstein statistics,
$$
N = \sum_{n} N_n^\pm, 
\qquad
N_n^\pm 
= {1\over{e^{\alpha^\pm + b e^\pm_n} - 1}},
\eqn\popl
$$
where we have
introduced 
$b = \beta {\cal E}$ ($\beta = 1/k T$).
Note that
$\alpha^\pm$ are determined by 
the particle number constraint (the first of (\popl)) and 
are dependent on the temperature.  The forces (or pressure) acting on the partition
from the right and the left are then given by
$$
F^\pm = - \sum_n {{\partial E_n^\pm}\over{\partial
l}} N_n^\pm.
\eqn\express
$$
For simplicity, in what follows we use the dimensionless force and temperature
defined by
$$
f^\pm = {{l}\over{2\cal E}} F^\pm, \qquad 
t = {1\over{b}} = {k\over{{\cal E}}}T,
\eqn\dlq
$$
with which the net force on the partition becomes
$$
\Delta f = f^{-} - f^{+}, \qquad 
f^\pm = \sum_n 
e^\pm_n N^\pm_n. 
\eqn\nfp
$$
Our objective is to find how the net force $\Delta f$ behaves 
as a function of the temperature variable $t$.

\topinsert
\epsfxsize 7.0cm
\ifx\omitpictures N   \centerline{\epsfbox {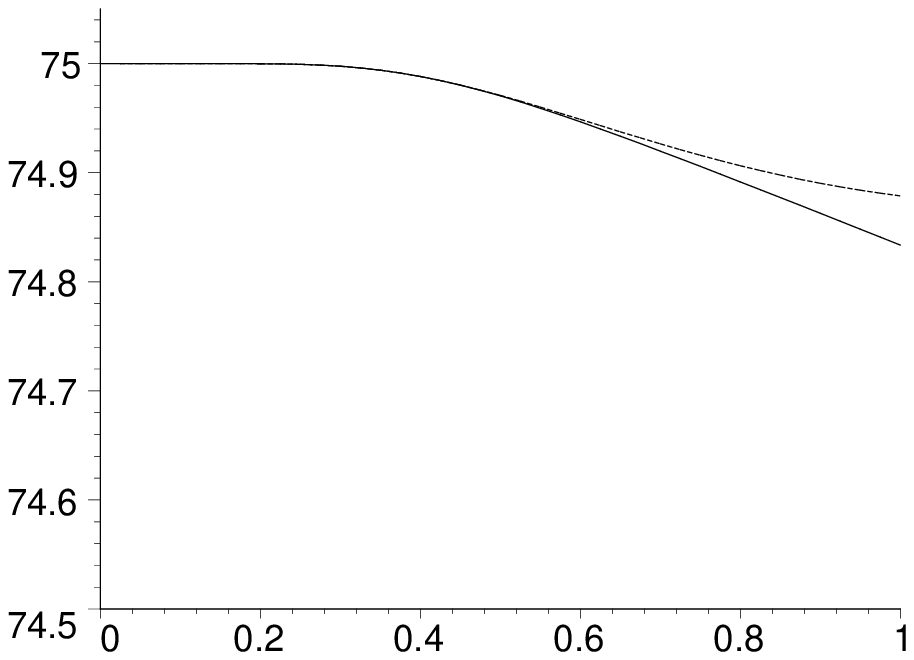}}  \fi
\yex{1}
\abstract{{\bf Figure 1.} The net force $\Delta f(t)$ for \mm{N = 100},
in the temperature region \mm{t < 1 \co} obtained by a numerical computation (solid
line), and approximated by Eq.~(2.8) (dashed line).}
\bigskip
\endinsert

To proceed, we first study the low temperature regime where a finite number of 
particles remain in the ground state $n = 1$.  This occurs if 
$\alpha^\pm + b e^\pm_1 \ll 1$, and in this regime
a numerical computation exhibits interesting
behaviors of the net force $\Delta f$.  Namely, as shown in Fig.1, $\Delta f$ starts
off by the value $\Delta f = 3N/4$ at $t = 0$ and decreases slightly but
basically stays there for 
$t < 1$.  Above $t \approx 1$ the net force starts to decrease almost linearly until
it reaches its minimum at around $t \sim N$ (see Fig.2), and from there it increases
to infinity.  To see how these behaviors arise, let us consider 
the case of the extremely low temperatures 
$t < 1$, where most of the $N$ particles are in the ground state.  
In this case, if we write $f^\pm$ in (\nfp) as
$$
f^\pm = e_1^\pm N + \sum_{n = 2}^\infty g_n^\pm,
\qquad g_n^\pm = \left(e_n^\pm - e_1^\pm\right) N^\pm_n,
\eqn\reexpress
$$
then we find that the first term gives the zero temperature value $f^\pm(0)$ 
and the rests represent the corrections from the higher energy levels at finite
temperatures.  Since $N^\pm_n$ decreases exponentially fast for higher $n$, we may keep
only the first contribution $n = 2$ to get
$$
\Delta f(t) \approx  {3\over 4} N + (3\, e^{-3/t} - 2\, e^{-2/t}).
\eqn\nfres
$$
This gives a good approximation for $t < 1$ as seen in Fig.1.

The linear decrease of the net force observed for temperatures higher (but not much
higher) than $t = 1$ may be understood heuristcially as follows.  First, we classify
the energy levels into three categories, where the first is those levels for which
$b(e_n^\pm - e_1^\pm) \ll 1$, the second is those for which 
$b(e_n^\pm - e_1^\pm) \gg 1$, and the third is the rest, {\it i.e.}, those for which
$b(e_n^\pm - e_1^\pm)$ is not far from 1.  Let
$m$ be the level whose $b(e_n^\pm - e_1^\pm)$ is closest to 1,
that is,
$b(e_m^\pm - e_1^\pm) \approx 1$.
Levels with $n \ll m$ thus belong to the first category, and for these one 
has $\alpha^\pm + b e_n^\pm \approx b(e_n^\pm - e_1^\pm)$ and hence 
$N_n^\pm \approx {1\over{b(e_n^\pm - e_1^\pm)}}$.
The corresponding $g_n^\pm$ given in (\reexpress) are then found to be
$g_n^\pm \approx {1\over{b}}$.  Thus we have $g_n^- - g_n^+ \approx 0$ showing
that the levels in this category do not contribute to the net
force.
On the other hand, levels with $n \gg m$ belong to the second category, and for
these we have
$N_n^\pm 
\approx e^{-b(e_n^\pm - e_1^\pm)}$ but
since
$b(e_n^\pm - e_1^\pm) \gg 1$, the corresponding force $F_n$ is seen to be
exponentially small and can be ignored.
Thus the contribution to the net force can comes only from the intermediate
levels belonging to the third category.  As a representative, let us 
choose
$m$ such that $b(e_m^+ - e_1^+) = 1$ holds in effect,
which is equivalent to 
$m^2 - m = t$, for the right half well.
Since we have $N_m^{+} = 1/(e-1)$,
we get 
$$
g_m^{+} =  {1\over b}\cdot{1\over{e - 1}}.
\eqn\qfmr
$$
In contrast, for the level $m$ for the left half well, we have
$b(e_m^- - e_1^-) = {{m^2 - 1}\over t} = 1 + {1\over m}$
and hence $N_m^{-} = 1/(e^{1 + 1/m} - 1)$.  Thus 
the contribution to the force becomes
$$
g_m^{-} = {1\over b}\cdot
{{1 + {1\over m}}\over{e^{1 + 1/m} - 1}}.
\eqn\qfml
$$
Combining (\qfmr) and (\qfml), 
and estimating roughly that 
the total number of levels in the third
category is of the order of $m$, ranging from $m/2 < n < 3m/2$, say, we just 
multiply the number $m$ to each of the contribution to find
$$
\Delta f(t) \approx  {3\over 4} N  + \sum_{n = m/2}^{3m/2} (g_n^- - g_n^+)
= {3\over 4} N - {t\over{(e - 1)^2}} .
\eqn\nflt
$$
The linear decrease of the net force is now seen in (\nflt).
Our argument assumes
$\alpha^\pm + b e_1^\pm < b(e_n^\pm - e_1^\pm)$ which gives the upper limit
of the temperature $t$ for which the heuristic formula (\nflt) is available.
The numerical result shows that the limit is around $t \approx 2N/3$.

\topinsert
\epsfxsize 7.0cm
\ifx\omitpictures N   \centerline{\epsfbox {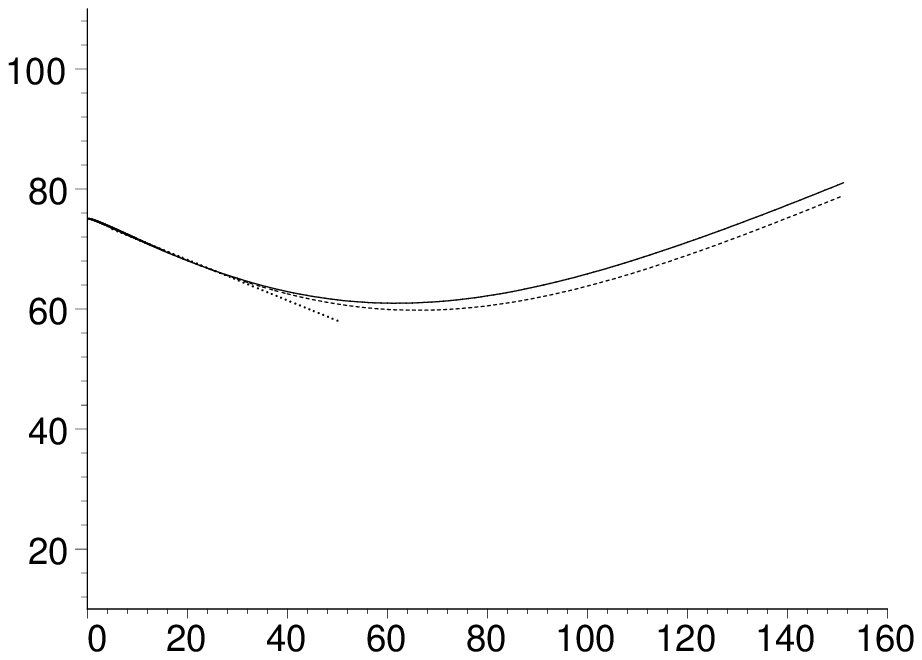}}  \fi
\yex{1}
\abstract{{\bf Figure 2.} The net force $\Delta f(t)$ for \mm{N = 100} and in the
temperature region \mm{ 0 < t < 160}, obtained by a numerical computation (solid line),
and approximated by Eq.~(2.11) (dotted line), and by Eq.~(2.16) using Eq.~(2.18)
(dashed line).}
\bigskip
\endinsert

If the temperatures are even higher $t \gg 1$ but still satisfy $\alpha^\pm + b
e^\pm_1
\ll 1$ remaining in the low temperature regime, the above approximation
becomes worse  and we need to resort to more systematic means to reproduce the
numerical results.  One such method valid for $t \gg 1$ is to consider the infinite
sums as the trapezoidal approximations of corresponding integrals. Thus we can write
$$
  \sum_{n=1}^{\infty} y(s_n) \approx \f{ y(s_1) }{2} +  
  \f{1}{\Delta s} \int_{s_1}^{\infty} y(s) \bit \d s
\eqn\apf
$$
for functions $y(s)$ vanishing at infinity, $\lim_{s \to \infty} y(s) = 0$. The
summation in (\apf) is taken at equidistant points,
$s_n$, $n = 1, 2, \ldots$, with $\Delta s = s_{n+1} - s_n =$ constant, and obviously 
the approximation is better for smaller $\Delta s$.
Using this approximation, we obtain
$$
\eqalign{
f^\pm 
&= \f{1}{b} \sum_{n=1}^{\infty}
  \f{ b e_n }{ \ee{ \alpha^\pm + b e^\pm_n } - 1 } 
= - \f{N \alpha^\pm}{b} + \f{1}{b} \sum_{n=1}^{\infty} 
  \f{ \alpha^\pm + (s^\pm_n)^2 }{ \ee{ \alpha^\pm + (s^\pm_n)^2 } - 1 } \cr
&\approx
   - \f{N \alpha^\pm}{b} + \f{1}{b} \[ \f{1}{2} 
  \f{ \alpha^\pm + (s^\pm_1)^2 }{ \ee{ \alpha^\pm + (s^\pm_1)^2 } - 1 } +
  \f{1}{\s{b}} \int_{s^\pm_1}^{\infty} 
  \f{ \alpha^\pm + s^2 }{ \ee{ \alpha^\pm + s^2 } - 1 } \d s \]
\co }
$$
where we have 
introduced $s^\pm_n = \s{ b e_n^\pm }$ which are equidistant 
on account of (\unitenergy).   Note that the increments 
\m{ \Delta s^\pm = \s{b} =  1/\s{t} } are indeed small for $t \gg 1$. 
Since \mm{ \f{z}{\ee{z} - 1} \approx 1 } for small \m{z}, 
we can approximate the second term by
\m{ \f{1}{2 b} }, and similarly the integral \mm{ \int_{0}^{s^\pm_1}
\f{ \alpha^\pm + s^2 }{ \ee{ \alpha^\pm + s^2 } - 1 } \d s } by
\m{s^\pm_1}. We thus obtain
$$
f^\pm \approx \f{ - N \alpha^\pm + 1/2 - s^\pm_1 / \s{b} }{b} +
  \f{1}{ b^{3/2} } \int_{0}^{\infty} 
  \f{ \alpha^\pm + s^2 }{ \ee{ \alpha^\pm + s^2 } - 1 } \d s \pe
\eqn\no
$$
Employing the formula 
$\f{z}{\ee{z} - 1} \approx \ee{-z} \( 1 + \f{z}{2} + \f{z^2}{12} \)$
which is handy for evaluating the integral approximately, we find 
$$
  \int_{0}^{\infty} \f{ \alpha^\pm + s^2 }{ \ee{ \alpha^\pm + s^2 } - 1 }
  \d s \approx 
\f{ \s{\pi} }{96} (63 - 35 \alpha^\pm) + \ordo{\alpha^\pm{}^2},
\eqn\no
$$
and hence
$$
f^\pm \approx 
  \( - N \alpha^\pm + 1/2 - \s{e^\pm_1} \) \qt +
  \f{ \s{\pi} }{96} (63 - 35 \alpha^\pm) \bitt \qt^{3/2} \pe
\eqn\no
$$
The net force is then 
$$
\Delta f \approx \( N \qt + \f{35}{96} \s{\pi} \bittt \qt^{3/2} \)
  \( \alpha\Neu - \alpha\Dir \)
  + \( \s{ e_1\Neu } - \s{ e_1\Dir } \) \qt \pe
\eqn\qnf
$$
In passing we remark that this formula (\qnf) turns out to be valid even for
small temperatures.  This can be seen, for example, by considering the \mm{t \to 0}
limit, where $\qN_1^\pm \approx N$ and hence from (\popl) we have
$\alpha^\pm \approx \ln( 1 + 1/N ) - b e^\pm_1$.  Thus the net force
(\qnf) is found to be $\Delta f \approx N \( e_1\Dir - e_1\Neu \) = 3N/4$
for $t \ll 1$ as seen in (\nfres).

To make use of (\qnf), we need to determine the \m{\alpha^\pm} as functions of
$t$ from the total number costraint in (\popl).  This can be done with the help of
the same approximation method as used above. Namely, upon using (\apf) 
we write down the constraint condition as
$$
\eqalign{
N 
&= \sum_{n=1}^{\infty} \qN_n^\pm = \qN_1^\pm 
+ \sum_{n=2}^\infty \f{1}{ \ee{ \alpha^\pm + b e_n^\pm } - 1 }\cr
&\approx \f{1}{ \ee{ \alpha^\pm + b e_1^\pm } - 1 } + 
  \f{1/2}{ \ee{ \alpha^\pm + b e_2^\pm } - 1 } + 
  \f{1}{\s{b}} \int_{s_2^\pm}^{\infty} 
  \f{ \d s }{ \ee{ \alpha^\pm + s^2 } - 1 },
  }
\eqn\no
$$
where we have kept \m{\qN_1^\pm} separately to achieve a better approximation (since
\m{\qN_n^\pm} are rapidly decreasing functions of \m{n} for small \m{n}).
Limiting the range of integration to \mm{ [0,\s{2}] } which provides the main
contribution to the integral, and using 
\mm{ \f{1}{\ee{z} - 1} \approx \f{1}{z} - \f{1}{2}} valid on the range, 
we find
$$
\eqalign{
N &\approx
\f{1}{ \alpha^\pm + b e^\pm_1 } + \f{1/2}{ \alpha^\pm + b e^\pm_2 } -
\f{3}{4} -\f{1}{ \s{b} } \f{ \s{2 - \alpha^\pm} - s^\pm_2 }{2} \cr
& \quad +\f{1}{ \s{b |\alpha^\pm|} }
 \[ \qA \( \f{\s{|\alpha^\pm|}}{s^\pm_2} \)
  - \qA \( \f{\s{|\alpha^\pm|}}{ \s{2 - \alpha^\pm} } \) \] \co }
\eqn\aat
$$
where the function \m{\qA} is the \mm{\arctan} function for positive
\m{\alpha^\pm}, and is the \mm{\hbox{arctanh}} function for negative
\m{\alpha^\pm}. Unfortunately, it is not easy to solve (\aat) directly for
\m{\alpha^\pm} even approximately to obtain an analytic formula
that can be used in (\qnf) to reproduce the numerical result.
Nevertheless, the formula (\aat) has allowed us to evaluate 
the infinite sum to a good accuracy, and we can solve (\aat) indirectly for
\m{\alpha^\pm} by numerical means.  The outcome of this semi-analytic analysis
is in good agreement with the numerical computation as shown in Fig.2.

\secno=3 \meqno=1

\bigskip
\noindent{\bf 3. High temperature regime}
\medskip

\topinsert
\epsfxsize 7.0cm
\ifx\omitpictures N   \centerline{\epsfbox {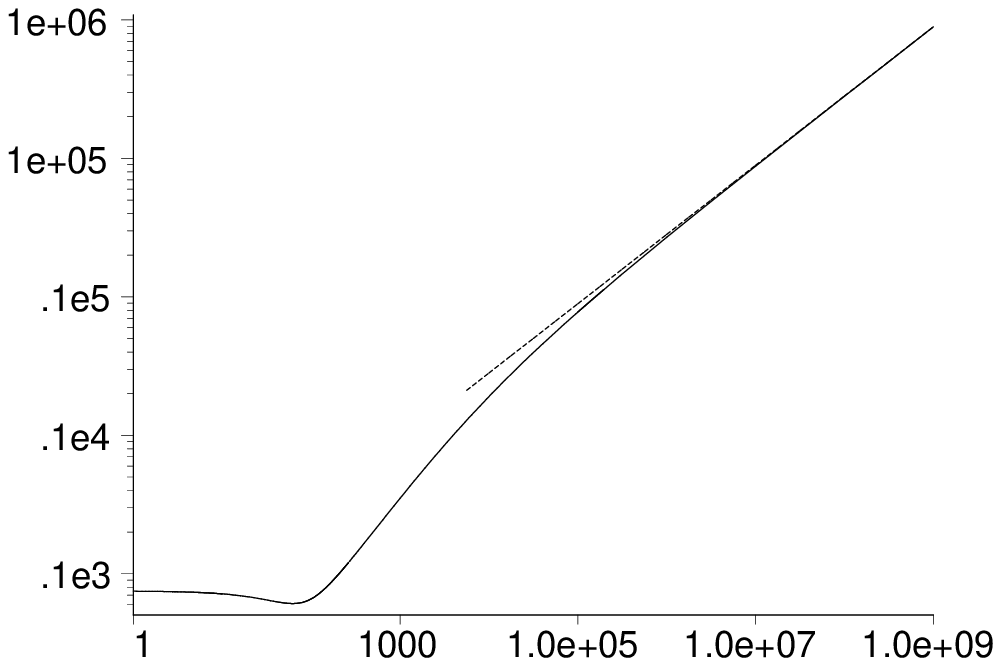}}  \fi
\yex{1}
\abstract{{\bf Figure 3.} The net force $\Delta f(t)$ for \mm{N = 100},
obtained by a numerical computation (solid line), and approximated for
high temperatures with Eq.~(3.7) (dashed line).}
\bigskip
\endinsert

Let us now determine the high-temperature asymptotic behavior of the net force.
For increasing temperature, we expect \mm{N^\pm_1} to decrease so [see
(\popl) for \m{n = 1}] we expect \m{\alpha^\pm} to increase to higher
positive values. Inspired by this, let us
expand \m{N_n^\pm} in \mm{ \qq^\pm := \ee{- \alpha^\pm} } as
  \E{\aad}{
  N_n^\pm = \f{ \qq^\pm \ee{- b e_n^\pm} }{ 1 - \qq^\pm \ee{- b e_n^\pm} }
  = \suml_{\qk = 1}^{\infty} (\qq^\pm)^\qk \ee{- \qk b e^\pm_n} \co
  }
which is valid for any positive \m{\alpha^\pm}. Thus
  \E{\aae}{
  N = \suml_{n = 1}^{\infty} N_n^\pm = \suml_{\qk = 1}^{\infty}
  (\qq^\pm)^\qk \suml_{n = 1}^{\infty} \ee{- \qk b e_n^\pm}
  = \suml_{\qk = 1}^{\infty} (\qq^\pm)^\qk \[ - \f{\qsigma^\pm}{2}
  + \f{1}{2} \suml_{n = - \infty}^{\infty} \ee{- \qk b e^\pm_n} \] ,
  }
with the constants \mm{\qsigma\Neu = 0 \co} \mm{\qsigma\Dir = 1} corresponding to the
\m{\pm} half wells, where we have extended the meaning of the notation \m{e_n^\pm} [see
(\unitenergy)] to negative \m{n}s, too. Applying now the Poisson summation formula
  \E{\aal}{
  \suml_{n = - \infty}^{\infty} \qf(n) = \suml_{\qm = - \infty}^{\infty}
  \intl_{-\infty}^{\infty} \d s \bitt \qf(s) \bitt \ee{ 2 \pi i \qm s } \co
  }
we obtain
  \E{\aag}{
  N = \suml_{\qk = 1}^{\infty} (\qq^\pm)^\qk \[
  - \f{\qsigma^\pm}{2} + \s{ \f{\pi}{ 4 \qk b } } \suml_{\qm =
  - \infty}^{\infty} (\qs^\pm)^m \ee{ - \f{\pi^2}{ \qk b } \qm^2 } \]
  }
with \mm{\qs^\pm = \mp 1 \pe}
Similarly, for the force \m{f^\pm}, one can find
  \E{\aah}{
  f^\pm = \suml_{n = 1}^{\infty} e_n^\pm N_n^\pm =
  \suml_{\qk = 1}^{\infty} (\qq^\pm)^\qk
  \s{ \f{\pi}{ 16 \qk^3 b^3 } } \suml_{\qm = - \infty}^{\infty}
  (\qs^\pm)^m \( 1 - \ff{2 \pi^2}{\qk b} \qm^2 \)
  \ee{ - \f{\pi^2}{ \qk b } \qm^2 } \pe
  }

For the high-temperature asymptotic behavior (\mm{\qq^\pm \to 0}), it
suffices to consider only the first some terms in the sums over
\mm{\qk} [both in (\aag) and (\aah)], and within each term to keep
only the \mm{\qm = 0} term in the sums over \m{\qm} (the \mm{\qm \ne 0}
terms being exponentially suppressed). Now, the leading, \mm{\qk = 1} term in (\aag)
gives that \mm{ \qq^\pm = 2 N \( \f{b}{\pi} \)^{1/2} + \ordo{b} \pe }
Since this leading behavior of
\m{\qq^\pm} is independent of \m{\qsigma^\pm}, inserting it into (\aah)
gives that the leading, \m{ \ordo{ b^{-1} } } term of \m{f^\pm} (coming from
\mm{\qk = 1}, \mm{\qm = 0}) is also \m{\qsigma}-independent. Hence,
this term will drop out from the net force. Therefore, to
have the first nonvanishing term in the net force we need
the first subleading term in \m{\qq}, too. Incorporating the \mm{\qk = 2} term
as well for \m{\qq^\pm}, we find
  $$
  \qq^\pm = 2 N \( \f{b}{\pi} \)^{1/2} + 2 N \[ \qsigma^\pm - \s{2} N \]
  \f{b}{\pi} + \ordo{ b^{3/2} } \pe
  \eqn\aai
$$
Plugging this into (\aah) and then calculating the net force yields
  \E{\aaj}{
  \Delta f = \f{N}{2} \( \f{\qt}{\pi} \)^{1/2} + \ordo{ \qt^{0} } \pe
  }
We can see in Fig.3 how the net force actually reaches this square-root asymptotic behavior
at high temperatures.

\secno=4 \meqno=1

\bigskip
\noindent{\bf 4. Discussion}
\medskip

We have found that the net force acting on the separating wall is nonzero at
low temperatures, being practically constant for very small temperatures and starting to
decrease when temperature is increased. Knowing that the energy spectrum is different on the
two halves this property is not very surprising. What is surprising, however, is that this
decrease stops at a certain temperature and the net force starts to increase above this
value. Furthermore, a remarkable fact is that this increase does not stop 
nor converges to some finite high-temperature limit but increases to infinity, as the square root
of the temperature. {}From the naive expectation that such quantum effects coming from the
different boundary conditions should vanish at high temperatures where the classical picture
would be available, this result seems quite unusual. However, this may be understood by the
fact that, contrary to most quantum systems, one dimensional boxes have such energy spectra that
the level spacing is not decreasing but increasing for higher energy levels (which is actually
valid not only for boxes with Dirichlet and/or Neumann boundary conditions but for all other boxes
as well [\FTC]). In other words, quantum boxes can be
distinguished by their high-temperature behavior, too.

We mention that the calculation presented here could be repeated for boxes with other
boundary conditions, too. We note however that, for most box systems, the energy values
are determined by a transcendental equation [\FTC], and hence a certain additional difficulty
for carrying out the calculations, especially the analytical ones, will arise there.

\bigskip
\noindent
{\bf Acknowledgement:}
This work has been supported in part by
the Grant-in-Aid for Scientific 
Research on Priority Areas (No.~13135206) by
the Japanese
Ministry of
Education, Science, Sports and Culture.

\baselineskip= 15.5pt plus 1pt minus 1pt
\parskip=5pt plus 1pt minus 1pt
\tolerance 8000
\yex{7}\immediate\closeout\reffile
\centerline{{\bf References}}\bigskip\frenchspacing%
\input refs.tmp\vfill\eject\nonfrenchspacing

\bye